\documentclass[conference]{IEEEtran}
\IEEEoverridecommandlockouts

\usepackage{cite}
\usepackage{amsmath,amssymb,amsfonts}
\usepackage{algorithmic}
\usepackage{graphicx}
\usepackage{textcomp}
\usepackage{xcolor}
\usepackage{subfigure}
\usepackage{url}
\usepackage{booktabs}
\usepackage{threeparttable}
\usepackage[framemethod=tikz]{mdframed}
\usepackage[normalem]{ulem}

\def\BibTeX{{\rm B\kern-.05em{\sc i\kern-.025em b}\kern-.08em
    T\kern-.1667em\lower.7ex\hbox{E}\kern-.125emX}}

\newcommand{\system}{\textsc{Pasta}}

\definecolor{insertcolor}{RGB}{56,132,28}
\definecolor{deletecolor}{RGB}{232,52,47}

\mdfdefinestyle{KeyFinding}{
    backgroundcolor=gray!20, 
    roundcorner=5pt,
}

\begin{document}

\title{Exploring~Direct~Instruction~and~Summary-Mediated Prompting in LLM-Assisted Code Modification}

\author{Ningzhi Tang\IEEEauthorrefmark{1}, Emory Smith\IEEEauthorrefmark{1}, Yu Huang\IEEEauthorrefmark{2}, Collin McMillan\IEEEauthorrefmark{1}, Toby Jia-Jun Li\IEEEauthorrefmark{1} \\
\IEEEauthorrefmark{1}\{ntang, esmith36, cmc, toby.j.li\}@nd.edu, \IEEEauthorrefmark{2}yu.huang@vanderbilt.edu \\
\IEEEauthorrefmark{1}University of Notre Dame, Notre Dame, IN, USA
\IEEEauthorrefmark{2}Vanderbilt University, Nashville, TN, USA
}

\maketitle

\begin{abstract}
This paper presents a study of using large language models (LLMs) in modifying existing code. While LLMs for generating code have been widely studied, their role in code modification remains less understood. Although ``prompting'' serves as the primary interface for developers to communicate intents to LLMs, constructing effective prompts for code modification introduces challenges different from generation. Prior work suggests that natural language summaries may help scaffold this process, yet such approaches have been validated primarily in narrow domains like SQL rewriting. This study investigates two prompting strategies for LLM-assisted code modification: Direct Instruction Prompting, where developers describe changes explicitly in free-form language, and Summary-Mediated Prompting, where changes are made by editing the generated summaries of the code. We conducted an exploratory study with 15 developers who completed modification tasks using both techniques across multiple scenarios. Our findings suggest that developers followed an iterative workflow: understanding the code, localizing the edit, and validating outputs through execution or semantic reasoning. Each prompting strategy presented trade-offs: direct instruction prompting was more flexible and easier to specify, while summary-mediated prompting supported comprehension, prompt scaffolding, and control. Developers' choice of strategy was shaped by task goals and context, including urgency, maintainability, learning intent, and code familiarity. These findings highlight the need for more usable prompt interactions, including adjustable summary granularity, reliable summary-code traceability, and consistency in generated summaries.
\end{abstract}


\section{Introduction}

The rise of large language models (LLMs) has transformed how developers interact with code~\cite{liang2024large, tang2024developer}. While much attention has been given to how LLMs generate new code implementations from natural language specifications~\cite{chen2021evaluating, vaithilingam2022expectation}, their role in \textit{modifying existing code}, a common yet cognitively demanding task, remains comparatively underexplored. Unlike code generation, modification requires developers to interpret existing logic, align their intent with that logic, and ensure that changes do not introduce regressions~\cite{gilmore1991models, lehman1985software, britton2013reversible}. Despite complexity, recent evidence shows that editing code via natural language prompts accounts for a substantial share (831/4,188) of code-related LLM usage~\cite{zheng2023lmsys, cassano2023can}, and this functionality is increasingly supported by production tools like GitHub Copilot Chat\footnote{https://github.com/features/copilot} and Cursor\footnote{https://www.cursor.com/}. This shift also reflects a broader trend toward ``vibe coding,'' where developers delegate code modifications to AI without making manual edits\footnote{https://x.com/karpathy/status/1886192184808149383}.

In such workflows, prompts serve as the primary interface through which developers communicate intent to LLMs~\cite{brown2020language, liu2023pre}. However, crafting an effective prompt remains difficult due to the inherent open-endedness of natural language~\cite{tankelevitch2024metacognitive, liu2023wants}, uncertainties in LLM predictions~\cite{chen2022machine, sarkar2022like}, and the often vague or evolving nature of developer intentions~\cite{zamfirescu2023johnny, tankelevitch2024metacognitive}. 
These challenges become especially prominent in code modification~\cite{liang2024prompts}, where the prompt must navigate complex, pre-existing code contexts and articulate precise changes. Developers struggle to understand unfamiliar logic, locate modification points, and express intent clearly~\cite{ko2006exploratory}, while LLMs must respond accurately without introducing side effects~\cite{chen2022machine}.

One promising approach to easing prompt construction is to scaffold it through editable natural language summaries of existing code, which serve as intermediate representations. Previous work in constrained domains, such as SQL rewriting~\cite{tian2023interactive, tian2024sqlucid} and spreadsheet data analysis~\cite{liu2023wants, ferdowsi2023coldeco}, has explored this strategy. They show that the summaries help bridge the gap between developer intent and LLM outputs, while also improving code comprehension and giving users more control over prompt formulation. However, these approaches have largely been limited to domains with small-scale code and semantically constrained programming languages, where summary generation is achieved through predefined rule-based mappings from code to linearized instructions. Their application to broader general-purpose programming scenarios remains underexplored. In these scenarios, both code semantics and developer modification intentions are more diverse and less structured. Additionally, general-purpose code often involves abstract control flow and cross-cutting dependencies, making rule-based summaries poorly aligned with human reasoning and task-specific goals.

\begin{figure*}[htbp]
    \centerline{\includegraphics[width=0.8\textwidth]{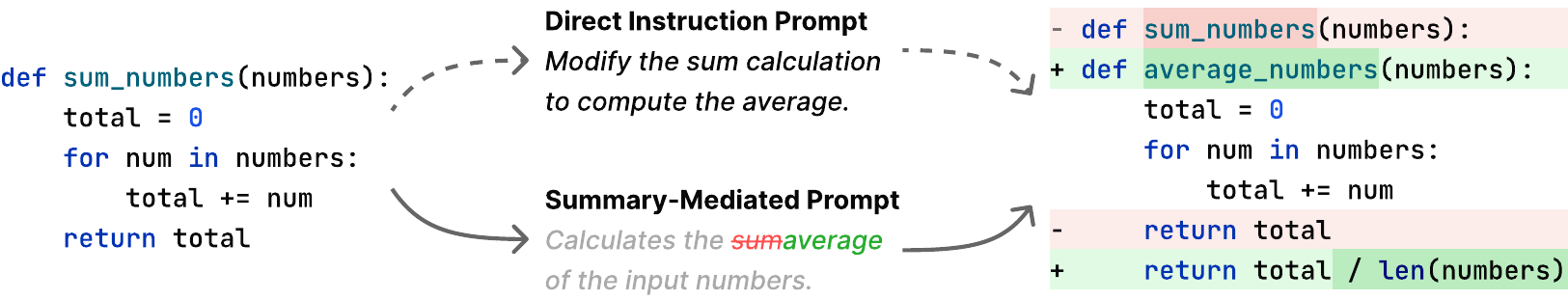}}
    \caption{An illustrative example of direct instruction (dashed arrows) and summary-mediated prompting (solid arrows) for LLM-assisted code modification.}
    \label{fig:illustrative_example}
\end{figure*}

To bridge this gap, our work investigates how developers use two prompting techniques in \textit{general-purpose programming} to understand the processes, trade-offs, and design implications of LLM-assisted code modification. This is enabled by LLMs' ability to generate editable summaries from \textit{arbitrary code}. As illustrated in Fig.~\ref{fig:illustrative_example}, the two techniques are:
\begin{itemize}
    \item \textbf{Direct Instruction Prompting:} Freely writing natural language instructions to describe desired changes.
    \item \textbf{Summary-Mediated Prompting:} Editing a natural language summary of existing code to specify new behavior.
\end{itemize}
We conducted an exploratory study with 15 developers (10 graduate students and 5 professionals), who performed code modification tasks across three scenarios, each containing multiple sub-requirements. To facilitate the study, we developed a prototype named \system{} (\textbf{P}rompt-\textbf{A}ssisted \textbf{S}oftware \textbf{T}r\textbf{A}nsformation), integrated into JetBrains IDEs, which supports both prompting techniques. We collected interaction logs for each session. Finally, we conducted semi-structured interviews in which participants reflected on their experiences.

In each task, we asked participants to switch freely between prompting approaches and develop hybrid strategies. Unlike controlled comparisons that focus narrowly on prompting effectiveness, our open-ended design enables a richer understanding of developer behavior. This design allows us to investigate not only \textit{which} prompting methods developers prefer, but also \textit{how} and \textit{why} they adapt their strategies to different task contexts, and \textit{what} usability improvements could make summary-mediated prompting more effective in practice. Our study investigates the following research questions:
\begin{itemize}
\item \textbf{RQ1}: What are developers' prompting processes and perceived characteristics of the two prompting techniques?
\item \textbf{RQ2}: What task goals and contextual factors influence developers' prompting choices?
\item \textbf{RQ3}: What usability considerations and design opportunities arise in summary-mediated prompting?
\end{itemize}

We highlight the following findings.
\begin{enumerate}
    \item Developers used generated summaries to understand code, narrow down modifications, and evaluate outcomes through execution or semantic reasoning. When specifying changes, they navigated trade-offs: direct instruction prompting offered flexibility and simplicity, while summary-mediated prompting provided precise terminology, broader context, and control over untouched code. Despite requiring more reading, summaries reduced typing and forced deeper understanding.
    \item Prompting strategies varied depending on task goals and contextual factors: developers favored direct instructions when changes were urgent, simple, large in scope, or clearly defined, and used summaries to support comprehension and control when aiming for long-term maintainability, learning, or bug avoidance, or when working with unfamiliar or complex code. These needs for deeper understanding and careful validation were amplified in industrial settings, where codebases are large, interdependent, and beyond LLM knowledge.
    \item The usability of summary-mediated prompting depended on granularity and alignment with developers' goals; participants called for structured formats, flexible detail levels, summary-code mappings, and consistent outputs to support efficient comprehension and accurate edits.
\end{enumerate}

\section{Related Work}

\subsection{LLM-Assisted Code Modification}

Code modification accounts for a substantial portion of software development effort~\cite{lehman1985software, britton2013reversible}. Researchers have focused on automating specific modification tasks using LLMs, including bug fixing~\cite{zhang2023self, chen2023teaching, jin2023inferfix, wei2023copiloting} and refactoring~\cite{cordeiro2024empirical}. For instance, Xia~\emph{et al.}~\cite{xia2024automated} improved patch generation in program repair by integrating real-time feedback into LLMs. In more general settings, datasets like InstructCoder~\cite{li2023instructcoder} and \textsc{CanItEdit}~\cite{cassano2023can} framed LLM code modification as mapping from \textit{(original code + instruction)} to \textit{modified code}. Cassano~\textit{et al.}~\cite{cassano2023can} found that descriptive instructions consistently outperform vague ones, but impose a high developer workload. Shi~\textit{et al.}~\cite{shi2024natural} similarly demonstrated the technical feasibility of using LLM-generated summaries for synchronized code editing, focusing on summary quality in surface-level editing cases.

While these works highlight what LLMs can accomplish for code modification, they pay limited attention to how developers actually formulate prompts in practice. In production tools like Cursor, prompting typically occurs through inline code selection or chat-based interfaces. A growing trend, popularized by Karpathy as ``vibe coding,'' involves describing modifications in natural language and delegating implementation details to LLMs, often without manually reviewing the resulting code. Our study fills this gap by examining developers' prompting strategies in such workflows.

\subsection{Developers' Interaction with LLMs}

User-centered studies have extensively examined how developers interact with LLM-powered tools and their perceived effectiveness~\cite{vaithilingam2022expectation, barke2023grounded, tang2024developer, tang2024codegrits, peng2023impact, liu2023refining, liang2024large, mozannar2024reading}. LLMs have been shown to improve productivity~\cite{ziegler2022productivity, peng2023impact}, but the code they generate often contains functionality issues~\cite{liu2023refining, liang2024large}, requiring significant developer effort to verify and repair~\cite{barke2023grounded, tang2024developer}, accounting for up to 38\% of developers' time~\cite{mozannar2024reading}. Despite the prevalence of code modification in LLM workflows~\cite{cassano2023can} and its support in production tools, few studies have examined how developers construct and adapt prompts in these contexts.

Non-expert developers face additional barriers, including limited vocabulary for prompting and difficulty understanding generated code~\cite{feldman2024non, nguyen2024beginning, dakhel2023github}. Modifying unfamiliar or legacy code imposes additional cognitive overhead, as developers must first comprehend existing code before making changes~\cite{von1995program, balz2010continuous}. To investigate these interactions, our study focuses on code editing tasks involving \textit{technically unfamiliar frameworks} (e.g., TensorFlow, D3.js) in realistic programming scenarios.

Sarkar~\textit{et al.}\cite{sarkar2022like} characterize the difficulty of iteratively aligning user intent with LLM outputs as an ``abstraction matching'' problem~\cite{sarkar2022like, liu2023wants}. This echoes Don Norman's ``gulf of execution''~\cite{hutchins1985direct}: getting the system to do what the user intends. Several systems have addressed this gap in code generation through structured interaction design~\cite{yen2024coladder, kazemitabaar2024improving, xie2024waitgpt, tang2024towards, masson2024directgpt, di2025enhancing}. They target new code generation rather than code modification, which requires interpreting existing logic and making iterative changes. We fill this gap by examining interactive prompting for code transformation.

\subsection{Prompting Strategies for LLM Programming}

Many LLM-based programming systems rely on natural language prompting~\cite{brown2020language, jiang2022discovering, mcnutt2023design}, but prompting remains cognitively demanding, especially for non-experts~\cite{dang2022prompt, chen2022machine, dang2023choice, xu2022ide, liang2024large}. Users often struggle with getting started~\cite{zamfirescu2023johnny} and expend significant metacognitive effort throughout the process~\cite{tankelevitch2024metacognitive, liang2024prompts}, including decomposing tasks and adjusting prompting strategies over time. Various prompting techniques have been proposed to better align LLM outputs with human intent~\cite{liu2023pre, wei2022chain, bach2022promptsource, marvin2023prompt}. For instance, few-shot prompting~\cite{brown2020language} and chain-of-thought reasoning~\cite{wei2022chain} enhance performance by providing contextual examples and intermediate reasoning steps. In programming contexts, developers use tactics such as giving examples, stating coding goals, or iterating through multi-turn conversations~\cite{denny2023conversing, wang2024enhancing, liang2024prompts}.

Recent studies have explored using \textit{intermediate representations} to scaffold prompting, such as sketches~\cite{yen2024code}, visual data operations~\cite{masson2024directgpt}, and editable natural language summaries. Liu~\textit{et al.}~\cite{liu2023wants} proposed ``grounded abstraction matching,'' translating Pandas code into natural-language utterances to support prompt refinement. Similarly, Tian~\textit{et al.}~\cite{tian2023interactive, tian2024sqlucid} used stepwise SQL explanations to help users identify and correct errors. These approaches improved task accuracy and user confidence but were limited to rule-based mappings in constrained domains. Rawal~\textit{et al.}~\cite{rawal2024hints} further showed that natural-language versions of code improved debugging performance in algorithmic tasks, suggesting the broader potential of summaries to support modification. However, it remains unclear how well these techniques apply to general-purpose programming, where code semantics and developer intent are more complex, motivating us to investigate how they function in real-world development contexts.

\section{Study Design}

We conducted an exploratory study\footnote{The study protocol has been approved by the IRB at our institution.} to understand how developers modify existing code using two prompting techniques: \textit{Direct Instruction Prompting}, where users write natural language commands, and \textit{Summary-Mediated Prompting}, where they edit LLM-generated summaries. Rather than running a controlled comparison, we aimed to observe how participants naturally adopt, combine, and adapt these strategies in realistic, iterative coding scenarios.

\subsection{Programming Tasks}

We created three tasks across diverse domains (deep learning, data visualization, and web development), implemented using Python, JavaScript, HTML, and CSS. Each task involved a familiar domain to the participants but incorporated programming techniques that they are less familiar with, allowing us to examine how technical familiarity influences prompting behavior~\cite{feldman2024non}.
Task design followed a structured, goal-driven process. One author first defined the domain scope and target technique. The programming scenarios were iteratively refined by the research team to balance realism and controllable complexity. Each task had a baseline implementation, followed by exploratory ideation to identify plausible enhancements in algorithm design, UI, or interaction.
We selected final tasks based on: (1) completion within 20 minutes; and (2) coverage across diverse domains, code structures, and modification types, including both logic- and interface-level changes.

\begin{itemize}
    \item \textbf{Task 1.} \textit{TensorFlow Autoencoder (deep learning)}: Modify a TensorFlow-based sparse autoencoder to update its model architecture, loss function, and learning schedule.
    \item \textbf{Task 2.} \textit{D3.js LineGraph (data visualization)}: Enhance the X-axis markers and point labels of a D3.js line graph visualizing skin cancer detection accuracy.
    \item \textbf{Task 3.} \textit{Chrome Extension Translator (web development)}: Improve the button effects, UI style, and front-end to back-end communication of a Chrome extension that integrates OpenAI’s API to perform machine translations.
\end{itemize}

\subsection{Participants}  

We recruited 15 participants (10 male, 5 female; ages 22-30, \textit{M} = 25.1, \textit{SD} = 2.46) through purposive sampling~\cite{etikan2016comparison}. All had prior experience with Python and JavaScript and were majoring in computer science or relevant engineering fields. The group included 10 graduate students and 5 professional developers, with an average of 6.94 years of programming experience. Each received a \$66 Amazon gift card as compensation for their time. All participants had used LLM tools for programming and code modification tasks (e.g., bug fixing and refactoring), such as ChatGPT (12), GitHub Copilot (11), Claude (9), DeepSeek (6), and Cursor (4). 

Most participants had general domain experience across the three task scenarios (13/15, 15/15, and 14/15), but technical familiarity varied: for TensorFlow, 3 had never used it, 9 were unfamiliar, and 3 were proficient; for D3.js, 9 had never used it, 5 were unfamiliar, and 1 was proficient; for Chrome Extension, 5 had never tried, 7 were unfamiliar, and 3 were proficient. A summary of each participant's demographics and technical background is presented in Table~\ref{tab:demographics}.

\subsection{\system: Study-Enabling Prototype}~\label{sec:study_system}
\looseness=-1
To investigate how developers use two prompting techniques in real-world code modification, we developed \system{}\footnote{Code available at: \url{https://github.com/TTangNingzhi/PASTA}}, a JetBrains IDE plugin that supports both approaches. 
\subsubsection{Interface}

\begin{figure*}[htbp]
    \centerline{\includegraphics[width=0.8\textwidth]{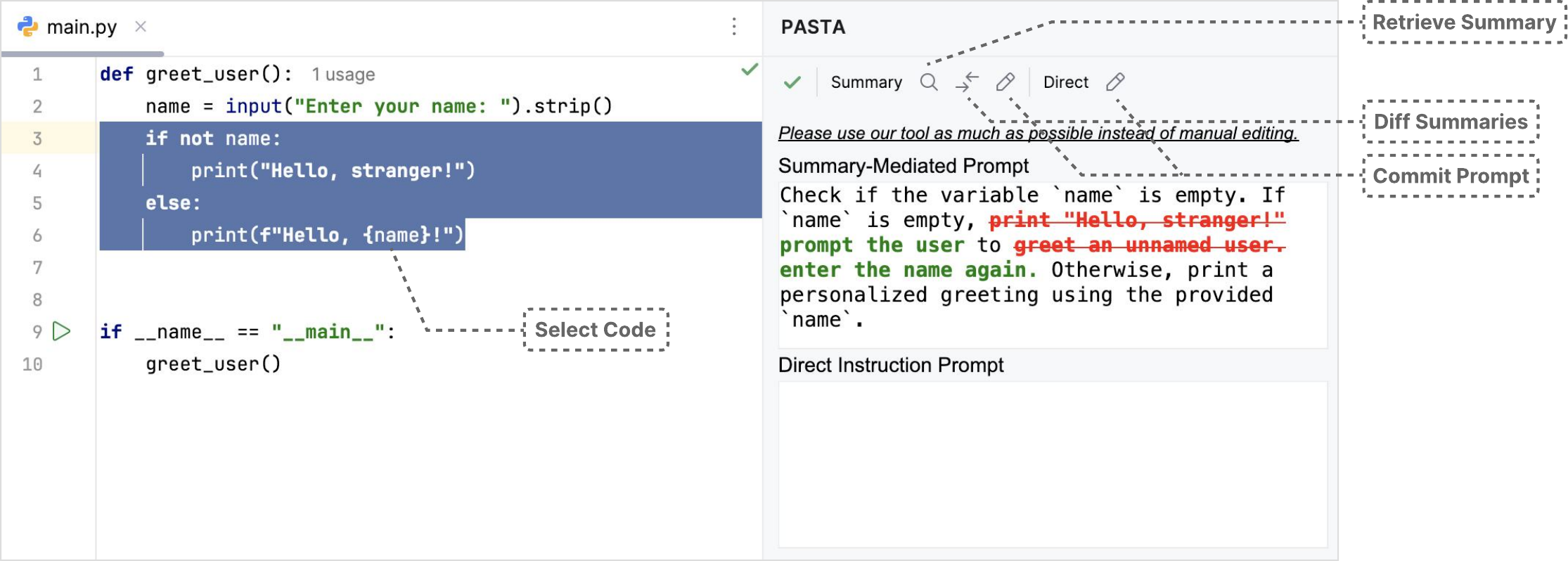}}
    \caption{Interface of the prototype plugin \system, integrated into JetBrains IDEs (e.g., PyCharm, WebStorm).}
    \label{fig:prototype_interface}
\end{figure*}

Fig.~\ref{fig:prototype_interface} shows the interface of \system{}. Developers begin by selecting a region of code in the editor using the mouse (left panel), then specify their intended modification using one of two prompting methods (right panel). 

In summary-mediated prompting, clicking~\raisebox{-0.2em}{\includegraphics[height=1em]{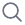}}~{\small\textsf{Retrieve Summary}} generates a 1-3 sentence natural language description of the selected code. The summary adapts to the length and complexity of the code. Pilot testing revealed that overly detailed summaries hindered usability, so we adopted a concise summarization policy. The generated summary appears in the top text field and can be freely edited to express the intended change. Developers can click~\raisebox{-0.2em}{\includegraphics[height=1em]{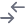}}~{\small\textsf{Diff Summaries}} to highlight {\small\texttt{\textcolor{insertcolor}{\textbf{insertions}}}} and {\small\texttt{\textcolor{deletecolor}{\textbf{deletions}}}}, aiding summary revision. In direct instruction prompting, developers write a natural language command in the bottom text field to specify the desired modification directly.

Clicking~\raisebox{-0.2em}{\includegraphics[height=1em]{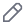}}~{\small\textsf{Commit Prompt}} in either mode submits the selected code, the file context, and either the edited summary or the free-form instruction to the LLM. The returned code is shown in a diff view with line- and token-level highlights, enabling developers to inspect, validate, and selectively accept changes. A loading indicator signals that the LLM is processing the prompt, providing users with timely visual feedback.

\subsubsection{Implementation}  

\system{} is implemented as a plugin using the official IntelliJ SDK\footnote{\url{https://plugins.jetbrains.com/docs/intellij/welcome.html}} and is compatible with all JetBrains IDEs, including PyCharm and WebStorm, which were used in our study. The interface is integrated as a tool window panel, with functional buttons implemented via the \texttt{AnAction} interface. The summary diff feature is supported by the Java Diff Utils library\footnote{\url{https://java-diff-utils.github.io/java-diff-utils/}}, while code diffing leverages the default diff package provided by the SDK. LLM-powered functionality is built on OpenAI's GPT-4o\footnote{\url{https://openai.com/index/hello-gpt-4o/}} chat completions API. To improve in-context learning and ensure output consistency, we include a set of few-shot examples in each prompt.

Unlike rule-based summary systems developed for SQL or tabular data~\cite{tian2024sqlucid, liu2023wants}, our approach leverages LLM generalization to describe arbitrary code without predefined templates. Although this sacrifices strict one-to-one correspondence between code and summary, it allows broader applicability across diverse programming domains.

\subsubsection{Design Decisions}\label{sec:design_decisions}

In designing \system{}, we made several key decisions to support effective prompting while maintaining compatibility with existing AI-assisted coding workflows.

\textbf{Selection-Based Prompting.}
Contemporary AI code editors, such as Cursor and GitHub Copilot Chat, typically offer two ways to specify target code for modification: (1) using the @ symbol in a chat interface to reference code, or (2) directly selecting code in the editor before entering a prompt. We adopt the latter selection-based approach for its simplicity, ease of integration, and natural alignment with summary generation, enabling a fair comparison between prompting techniques.

\textbf{Context-Aware Interaction.}
Prior work shows that incorporating relevant context, e.g., current file~\cite{su2024revisiting, haque2020improved}, call graphs~\cite{bansal2023function}, can improve LLM output for code tasks. Since our focus is not on comparing context strategies, we adopt an approach that includes the full content of the current file as context. This method is simple and effective for lightweight tasks, whereas how to effectively integrate broader context in large repositories remains a challenging and open question\footnote{\url{https://x.com/karpathy/status/1937902205765607626}}.

\subsection{Study Protocol}

\subsubsection{Settings}  

We conducted this study on a Windows 11 computer with PyCharm and WebStorm 2024.3 installed, along with our enabling research prototype, \system{} (Section~\ref{sec:study_system}). Participants accessed the computer via Zoom's remote screen control. We instructed participants to view the study instructions on a separate device (e.g., iPad or second screen) to save screen space and prevent direct copy-pasting into prompts.
During the study, we collected interaction logs from \system{}, including timestamps, selected code snippets, prompts, and LLM-generated summaries or modifications.

\subsubsection{Procedure}

Each study session lasted approximately 2 hours. We asked participants to sign a consent form and complete a pre-study questionnaire collecting their demographic information and prior experience with programming and LLM usage. We then provided a brief introduction to the study objectives, followed by a short tutorial on the two prompting ways through \system{}. 
Before starting the programming tasks, participants were given 5 minutes to try out both prompting methods on an example task.

The order of the three tasks was randomized to mitigate learning effects~\cite{nielsen1994usability}. For each task, participants first read an instructional document describing the background of the task and the required modifications. They were then given 20 minutes to complete each task. Participants were free to use either prompting technique; however, they were encouraged to use \system{} to prompt (in either technique) rather than editing manually. Participants were free to ask the study coordinator how to run the code and check the results. We recorded task completion and logged the time taken for each task.

After completing each task, participants filled out a NASA Task Load Index (NASA-TLX) questionnaire~\cite{hart1986nasa} to self-report their cognitive workload. NASA-TLX is a widely used subjective tool that measures the user's perceived workload when performing a task. It includes six dimensions: Mental Demand, Physical Demand, Temporal Demand, Performance, Effort, and Frustration. We also asked participants to self-report their code understanding and editing effort.

After completing all tasks, participants filled out a utility evaluation questionnaire with Likert scale items to assess their experience with each prompting technique, including their ease of use, effectiveness in code comprehension and modification, sense of control, and satisfaction with LLM-generated modifications. Finally, we conducted a 25-minute semi-structured interview of participants' experiences when using the two prompting techniques. The interview covered topics including prior experience with LLM-based code modification, challenges in prompting and validation, perceived control over LLM-generated changes, and preferences between the two prompting techniques. We also explored participants' views on the usability of generated summaries and potential improvements to summary-mediated prompting.

\subsection{Data Analysis}

We transcribed all post-study interviews and conducted a qualitative analysis following open coding procedures~\cite{brod2009qualitative, lazar2017research}. In the first stage, two authors collaboratively reviewed interview transcripts and extracted 174 meaning-rich segments related to developers' prompting strategies, decision-making processes, and challenges. These segments served as the units of analysis. Based on these segments, the authors developed an initial hierarchical codebook, consisting of high-level themes and nested subcodes. The codebook was iteratively refined through discussion. In the second stage, both authors independently coded all segments using the initial codebook. We assessed inter-coder reliability using both agreement rate and Cohen's $\kappa$, which yielded an agreement of 63.8\% and $\kappa$ = 0.626, indicating substantial consistency~\cite{landis1977measurement}. Disagreements were resolved through discussion. 

To analyze the interaction logs of the two prompting choices, we used Student's t-test for group comparisons with unequal sample sizes, reporting both the mean and $p$-value. For Likert-scale utility ratings, we applied the Wilcoxon signed-rank test, given the ordinal nature of the data and small sample size, and reported the median and $p$-value.

\section{Study Results}

\begin{figure}[tbp]
    \centerline{\includegraphics[width=0.5\textwidth]{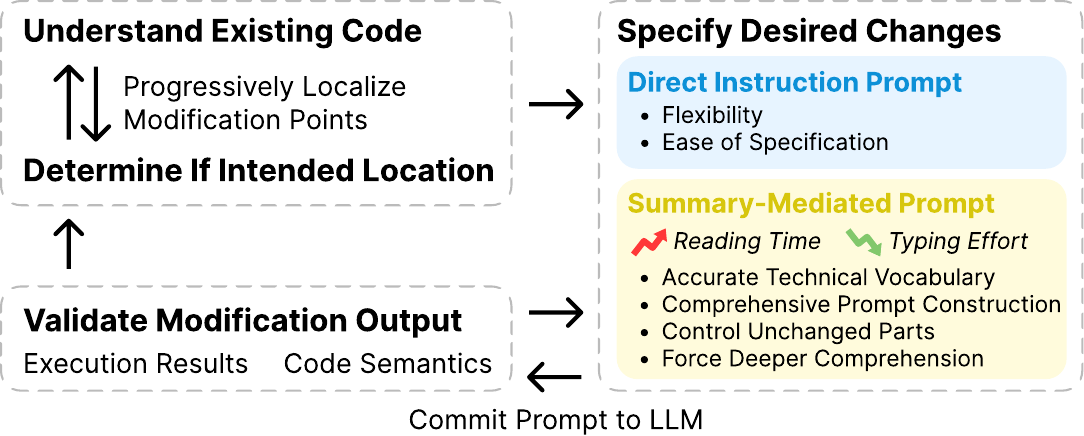}}
    \caption{A process model of developer-LLM interaction for code modification, highlighting complementary trade-offs between prompting strategies.}
    \label{fig:developer_process}
\end{figure}

\begin{figure*}[htbp]
    \centerline{\includegraphics[width=0.8\textwidth]{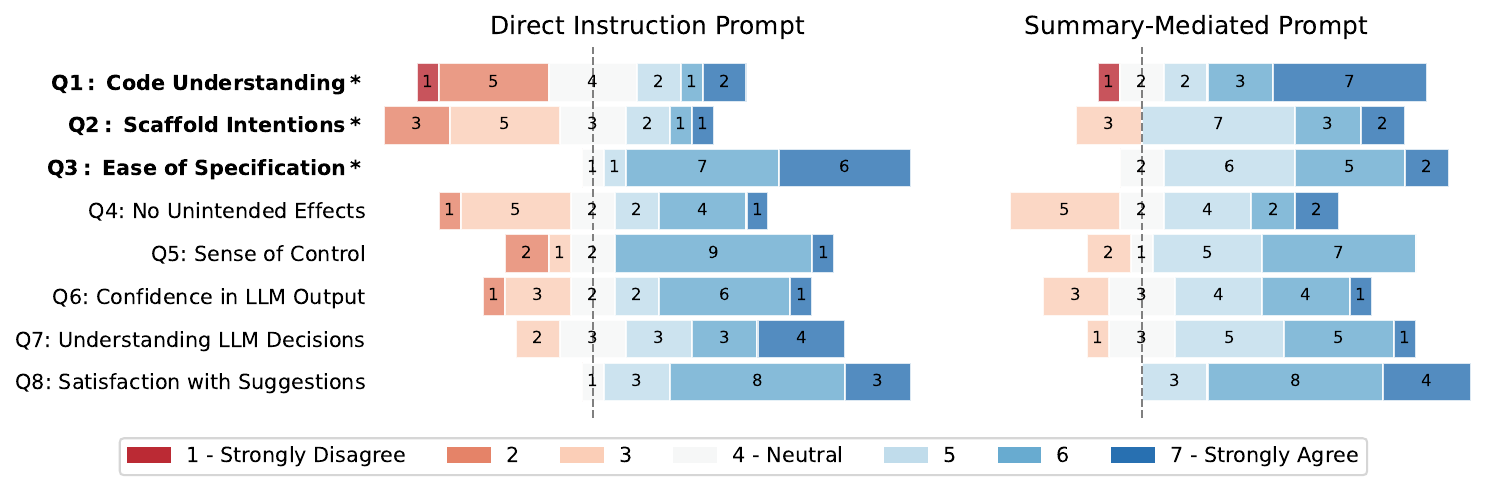}}
    \caption{Comparison of the usability of the two prompting techniques. Q1-Q3 were bolded with asterisks (*) to indicate statistically significant differences ($p < 0.05$), based on Wilcoxon signed-rank tests. No significant differences were observed for Q4-Q8.}
    \label{fig:likert_plot}
\end{figure*}

\subsection{Prompting Processes and Perceived Characteristics}

In this section, we present a process model of LLM-assisted code modification (Fig.~\ref{fig:developer_process}), highlighting how the two prompting strategies support different stages, from understanding and localization to specifying and validating changes, and analyze how summaries can first aid comprehension and later serve as a medium for constructing prompts during modification.

\subsubsection{Summaries Scaffold Code Understanding and Localization}

Before specifying changes, developers first need to understand the existing code and progressively localize the modification point (Fig.~\ref{fig:developer_process}), a process where summaries played a key role. As P5 summarized, ``\textit{Summary helped in two ways: understanding the code and deciding which part to modify.}''

\textbf{Summaries support comprehension of unfamiliar code (Q1).}
Participants found summaries helpful for understanding code that they hadn't authored themselves (P1, P2, P3, P5, P6, P8, P9, P12, P13), noting they ``\textit{saved much time compared to looking into the documentation.} (P15)'' P3 added, ``\textit{Through the natural language description, I can infer where the bug might be},'' echoing findings that natural language improves bug detection by making program logic more explicit~\cite{rawal2024hints}. Quantitative results also support this role: summaries were rated more helpful for understanding code (Fig.~\ref{fig:likert_plot}~Q1, $6.0 > 4.0$, $p = 0.0044 < 0.01$). Usage logs confirm frequent summary use, averaging 3.07, 4.00, and 5.33 invocations per task.

\textbf{Summaries enable developers to progressively localize the modification point and determine if it was the intended location} (P1, P3, P5, P8, P9, P14).
``\textit{Figuring out where to modify is challenging}'' (P14)~\cite{ko2006exploratory}, but summaries helped narrow the scope. As P9 described, ``\textit{I start by summarizing the entire file to understand its purpose, then gradually narrow the scope to a 50-line function that reveals internal logic}''. Similarly, P14 used the summaries to ``\textit{tell me what each line does}'' before locating the target section. As they refined the scope, participants continuously checked whether the selected code aligned with their editing intent. As P1 reflected, ``\textit{I might think, `Okay, this function is not the one that I should modify. I need to switch to another function and check.'}'' This progressive narrowing process echoes cognitive models of code navigation and bug localization~\cite{gilmore1991models, lawrance2010programmers}, where the iterative generation of summaries serves as a lightweight strategy to support hypothesis formulation and validation.

\textbf{Despite reading effort, developers frequently used summaries before choosing prompting strategies.}
Some participants (P1, P2, P13, P15) noted that long or detailed summaries slowed their workflow; P2 remarked, ``\textit{As a non-native English speaker, it took me some time to read the summarization response.}'' Structuring summaries and adding visual code mappings were suggested to improve comprehension (see Section~\ref{sec:design_opportunities}). Despite the overhead, many participants (P1, P2, P5, P8, P9, P12, P13, P14) reported a common pattern: they first used summaries to understand the code, narrow the scope, and then selected prompting strategies accordingly.

\subsubsection{Specifying Changes: Balancing Control, Effort, and Expressiveness}
\label{sec:specify_changes}

\begin{figure}[htbp]
    \centerline{\includegraphics[width=0.4\textwidth]{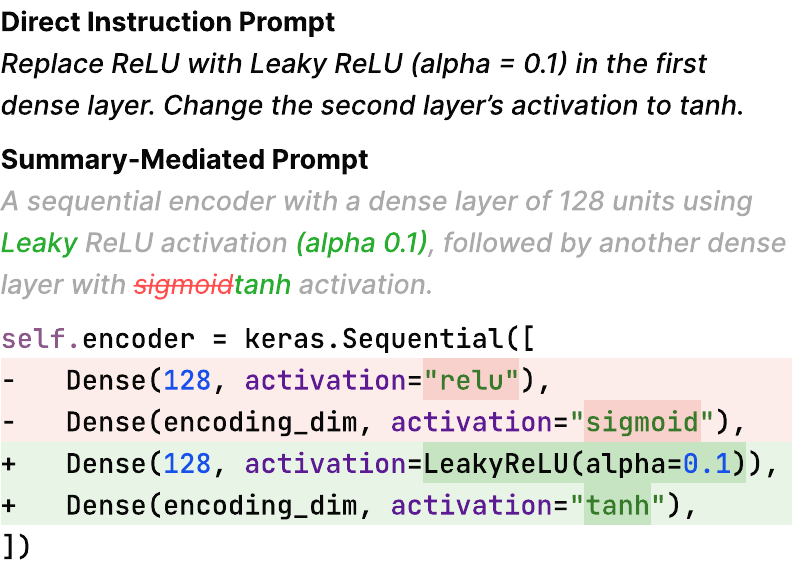}}
    \caption{Example prompts for fulfilling the first requirement in Task 1.}
    \label{fig:prompt_example}
\end{figure}

Once developers identified where to modify, they adopted two distinct strategies to specify the desired changes: direct instruction prompting (63.2\% usage), which offers ease and flexibility, and summary-mediated prompting (36.8\% usage), which provides scaffolding and finer control over the changes (Fig.~\ref{fig:developer_process}). 
Across all subtasks, 42.2\% used direct-only, 31.9\% summary-only, and 25.9\% mixed prompting, with respective success rates of 84.2\%, 93.0\%, and 80.0\%.

\textbf{Direct instruction prompting provided strong flexibility and ease of specification, allowing developers to quickly express desired changes (Q3).}
Many participants found direct instruction prompting ``\textit{intuitive and quicker}'' (P8), especially for straightforward edits. As P1 noted, ``\textit{I can directly write my command [...] and I'm more confident the modifications will align with my intentions, since it’s all my own input.}'' This flexibility allowed developers to specify changes rapidly without relying on generated summaries (P1, P2, P7, P12, P14). Participants rated direct instruction prompting higher in ``ease of specification'' (Fig.~\ref{fig:likert_plot}~Q3, $6.0 > 5.0$, $p = 0.0258 < 0.05$).  

However, this came with trade-offs, as developers acknowledged that direct instructions were more ``\textit{prone to errors}'' (P1) and could result in unintended modifications (P1, P8, P9). As P9 cautioned, ``\textit{if the model misunderstands my intent, it might touch unrelated parts of the code.}''

\textbf{Summary-mediated prompting provided accurate technical vocabulary, helping developers express changes more precisely.}
Using correct terminology can be challenging, especially for novice developers~\cite{nguyen2024beginning, dakhel2023github}. LLM-generated summaries ``\textit{use more accurate terminology, and then I can reuse those words}'' (P6), potentially reducing the effort of formulating instructions. As shown in Fig.~\ref{fig:prompt_example}, terms like {\small\texttt{activation}} and {\small\texttt{dense layer}} reflect domain knowledge that may not be easily recalled, especially by those less familiar with specific APIs or syntax (P6, P9, P13).

\begin{figure}[htbp]
    \setcounter{subfigure}{0}
    \subfigure[Distribution of Levenshtein distances (top 1\% outliers excluded).]{
        \includegraphics[width=0.53\columnwidth]{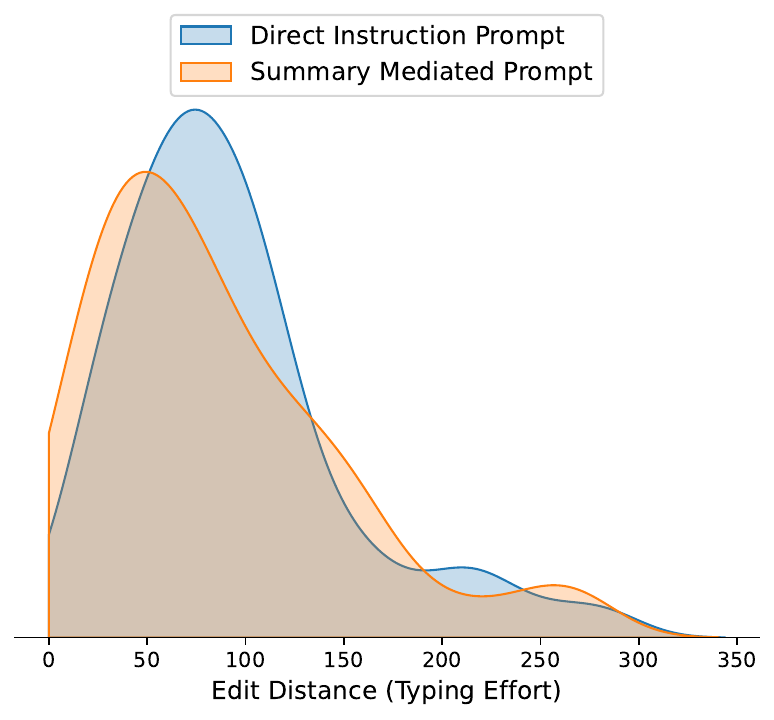}
        \label{fig:typing_distribution}
    }
    \hfill
    \subfigure[Box plot of selected code lengths (top 1\% excluded).]{
        \includegraphics[width=0.38\columnwidth]{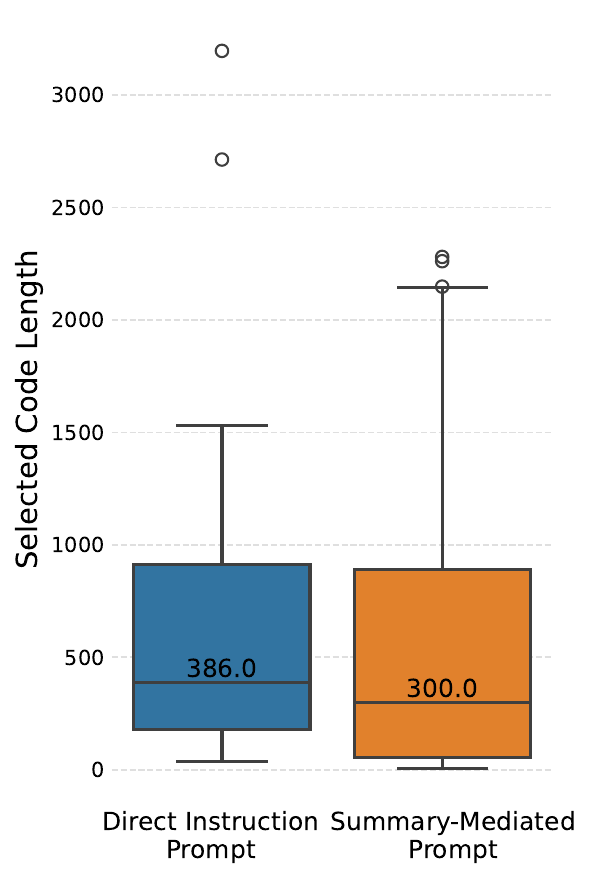}
        \label{fig:code_length_boxplot}
    }
    \caption{Comparison of interaction metrics. We did not observe a statistically significant difference.}
    \vspace{-2mm}
\end{figure}

\textbf{Editing summaries reduced typing effort by letting developers tweak existing text rather than writing prompts from scratch.}
Participants found this approach more natural and less effortful (P2, P3, P4, P6, P10, P12, P13). P3 shared, ``\textit{I prefer to [...] tweak the sentence directly, rather than inventing a full prompt from scratch.}'', as illustrated in Fig.~\ref{fig:prompt_example}. Fig.~\ref{fig:typing_distribution} shows that summary-mediated prompts required fewer keystrokes on average, measured by Levenshtein distance\footnote{\textbf{Levenshtein distance} measures the number of character-level insertions, deletions, or substitutions needed to transform the original summary (or an empty string for direct prompting) into the final prompt.}~\cite{levenshtein1966binary} ($82.2 < 89.7$), though not significantly ($p = 0.2248$). However, this benefit depended on the usability of the generated summaries (see Section~\ref{sec:usability}).

\textbf{Summaries served as scaffolds for composing comprehensive prompts, helping developers capture and express intended changes (Q2)} (P3, P4, P8, P10, P11, P13).
Participants rated summary-mediated prompting higher for supporting intent scaffolding (Fig.~\ref{fig:likert_plot}~Q2, $5.0 > 3.0$, $p = 0.015$). P8 noted that, compared to direct instruction, editing the summary helped ``\textit{preserve the semantic integrity of these encapsulated [function] units}.'' P11 also shared, ``\textit{Using the summary as a mental checklist seemed like a good idea [...] to cover more complex modifications}.'' Even when the original sentence was hard to modify directly, some participants preferred to leave it unchanged and add new sentences to extend its meaning (P3, P4, P6, P9, P10). This echoes prior findings that users build on existing utterances to guide system behavior~\cite{liu2023wants}.

\textbf{Summary-mediated prompting helped developers control unchanged parts of the code.}
While Q5 in Fig.~\ref{fig:likert_plot} showed no significant difference in perceived control between prompting techniques ($6.0 > 5.0$, $p = 0.893$), interviews revealed differing interpretations of ``control'': some focused on flexibility and speed, others on stability and scope. As illustrated in Fig.~\ref{fig:prompt_example}, summary-mediated prompting was helpful for constraining edits to intended areas (green/red text) and avoiding unintended changes (gray text). As P4 explained, ``\textit{If I change something small while keeping the rest [of the summary] the same, it helps constrain the LLM from modifying unintended areas.}'' P9 added that summaries ``\textit{retain an overview of the full code, including untouched parts, which gives me a stronger sense of safety.}'' This scoped control was especially valuable in complex tasks where developers needed to preserve existing semantics while making targeted changes.

\textbf{Editing summaries forced deeper comprehension of both existing code and intended modifications.}
Participants noted that reading and editing summaries required effort but led to stronger comprehension before making modifications (P2, P3, P9, P12). P3 admitted, ``\textit{Reading the summary takes time, [...] but I prefer to understand every line to modify with confidence.}'' This enforced comprehension clarified their intentions and improved control. As P2 reflected, ``\textit{I feel a much better sense of control because I have to fully comprehend to modify the code.}'' While summary-mediated prompting demanded more reading, it encouraged deliberate and accurate edits.

\subsubsection{Developers Balanced Quick Execution and Semantic Understanding When Validating LLM Modification Output}

Developers used two primary strategies: executing the code for immediate feedback or examining its semantics to avoid hidden issues, consistent with prior observations~\cite{o2025scientists}.

\textbf{Many prioritized quick execution to assess output correctness} (P1, P4, P7, P9, P10, P11, P12, P13, P14, P15).
P12 said, ``\textit{After I get the code, I just accept it and run it, then do my own debugging session.}'' This strategy was common when time was limited and participants trusted the LLM (P7).

\textbf{However, developers recognized that runtime checks alone could miss hidden errors~\cite{myers2011art}, leading them to prioritize semantic understanding.}
P14 warned, ``\textit{GPT adds or omits elements that you actually need, and these might not be immediately noticeable in the execution.}'' P3 believed that early comprehension reduces future debugging costs, noting, ``\textit{If I don’t understand the modified code and just accept it, the risk increases as the project grows.}'' For long-term reliability, participants stressed the need to understand the modified code. As P4 explained, summary-mediated prompting process forced them to ``\textit{understand what the code is actually doing.}''

These strategies were not chosen arbitrarily; developers' underlying goals, such as urgency, maintainability, or learning, shaped how they validated LLM outputs and, in turn, which prompting techniques they used. See Section~\ref{sec:intentional_factors} for details.

\begin{mdframed}[style=KeyFinding]
\textbf{Key findings:} Developers followed an iterative workflow: with the help of summaries, they first understood the existing code and progressively localized the modification point, then validated outputs through execution or semantic reasoning. When specifying changes, they balanced two prompting strategies: direct instruction prompting, valued for its flexibility and ease of specification, and summary-mediated prompting, which supported accurate terminology, contextual completeness, and control over unchanged code. Though summary-mediated prompts required more reading, they reduced typing effort and encouraged deeper comprehension.
\end{mdframed}

\subsection{Factors Shaping Prompting Strategy Choices}
\label{sec:factors}

Developers' prompting choices were shaped by multiple factors, which we categorize into two groups: (1) \textbf{intentional factors}, reflecting developers' underlying goals and motivations, and (2) \textbf{situational factors}, arising from the specific characteristics of the task and environment. We additionally draw on insights from professional developers to examine how these factors manifest in \textbf{industrial contexts}.

\subsubsection{Intentional Factors}
\label{sec:intentional_factors}

Intentional factors reflect developers' personal goals, shaping not only their prompting choices but also how they validated LLM-generated modifications.

\textbf{\uline{Task Urgency}. Urgent tasks led developers to prefer direct instruction prompting for its speed.}
Under time pressure or in one-off scenarios, developers prioritized fast completion over deep code comprehension, a pattern consistent with prior work on decision-making under time pressure~\cite{endsley1995toward}. As P1 put it, ``\textit{I just wanted to ensure the code works.}'' P8 similarly noted that in urgent situations with familiar code, they would ``\textit{skip the comprehension and focus on whether the execution is correct.}'' Direct instruction prompting enabled quick specification and execution-based validation, avoiding the overhead of reading or editing summaries.

\textbf{\uline{Codebase Maintenance}. Developers preferred summary-mediated prompting to ensure code quality in support of long-term maintenance interests.}
When working on personal or maintainable projects, developers prioritized understanding and control over changes. P13 emphasized, ``\textit{If it's my own project that requires long-term management, then a complete understanding is necessary.}'' Summary-mediated prompting enforced deeper comprehension, enabling safer changes aligned with long-term maintenance interests, an essential concern for code quality and modifiability~\cite{swanson1976dimensions, fowler2018refactoring}.

\textbf{\uline{Learning Goals}. When aiming to learn, developers used summary-mediated prompting to build understanding.}
Developers often treat unfamiliar tasks as learning opportunities and rely on summaries to explore code logic. P3 shared that summaries helped them ``\textit{learn something new in just a few seconds},'' while P7 noted that in real projects, they ``\textit{would take the time to learn and analyze the generated code more carefully}.'' Summary-mediated prompting supported learning by encouraging deeper understanding, aligning with concerns that effective GenAI use requires active cognitive effort to prevent shallow comprehension and reasoning offloading~\cite{singh2025protecting}.

\textbf{\uline{Bug Avoidance}. To avoid hidden bugs, developers favored summary-mediated prompting for early semantic validation.}
Participants aiming to avoid subtle errors were more cautious. As P14 warned, ``\textit{A small trick added by GPT can lead to a bug that is only discovered later.}'' P3 further explained, ``\textit{Debugging later may cost more than checking summaries early on.}'' P7 echoed this concern, ``\textit{When new errors appear, I have to go back, scrutinize my code, and debug it again, which can sometimes take longer.}'' By fostering early comprehension of code changes, summary-mediated prompting helped developers catch issues before they propagated.

\subsubsection{Situational Factors} \label{sec:situational_factors}

Beyond personal goals, task-related factors also shaped developers' prompting choices, influencing how they balanced efficiency and comprehension.

\textbf{\uline{Code Familiarity}. Familiarity with the codebase encouraged developers to favor direct instruction prompting.}
When familiar with the code, developers often skipped summaries and used direct instructions to specify changes efficiently because ``\textit{I knew exactly which part of the codebase I wanted to modify}'' (P5). In contrast, unfamiliar code prompted greater use of summary-mediated prompting to support understanding before edits (P5, P6, P8, P9, P10, P11, P12, P13), consistent with findings that comprehension strategies depend on prior experience and familiarity~\cite{letovsky1987cognitive, von1995industrial}. Spearman correlation analysis supports this trend, though correlations were modest. Developers with higher self-reported technical familiarity retrieved fewer summaries ($\rho = -0.15$), made fewer LLM-assisted modifications ($\rho = -0.12$), selected slightly larger code spans ($\rho = 0.06$), and used direct instruction prompts more frequently ($\rho = 0.05$). They also reported lower cognitive load, with mental demand negatively correlated ($\rho = -0.26$, $p = 0.084 < 0.10$). Other NASA-TLX dimensions (e.g., physical, temporal) and perceived performance followed similar trends. These patterns suggest that code familiarity reduces both reliance on external scaffolding and the perceived effort of LLM-assisted code modification.

\textbf{\uline{Task Difficulty}. Complex tasks lead developers toward using summary-mediated prompting to build initial comprehension.}
For simple, well-understood tasks, developers wrote direct instruction prompts ``\textit{because I trust the LLM can actually solve it}'' (P1, P2, P8). However, as task complexity increased, developers more often started with summaries to understand code structure and clarify requirements (P1, P2, P5, P8, P9, P12, P13, P14), reflecting the need for deeper comprehension under higher cognitive demand~\cite{letovsky1987cognitive}.

\textbf{\uline{Edit Scope}. Smaller edit scopes favored summary-mediated prompting for control; larger scopes favored direct instruction for flexibility.}
For small edits, developers valued the control summaries provided (P2). In contrast, broader changes led them to prefer direct instruction to avoid the overhead of reading or editing extensive summaries (P1, P2). This trend appears in Fig.~\ref{fig:code_length_boxplot}, where direct instruction prompts were linked to slightly longer code spans ($386.0 > 300.0$), though the difference was not statistically significant.

\textbf{\uline{Intention Clarity}. Clear goals led developers to prefer direct instruction prompting.}
When developers knew what they wanted to change, they expressed it directly and efficiently. As P11 noted, ``\textit{If I have a clear goal, I would go direct because I can just describe it, highlight the part, and tell it what to do.}'' In contrast, when intentions were vague, developers used summaries to help refine and clarify their objectives.

\subsubsection{Industrial Context}
\label{sec:industry_factors}

\textbf{The complexity and collaborative nature of industrial codebases increased developers' need for comprehensive understanding.}
Industrial projects span large, interdependent modules where developers focus on specific features while relying on shared infrastructure and APIs~\cite{von1995industrial}. Developers ``\textit{first need to understand how others' components work},'' (P8) and ``\textit{ensure that my AI modifications don't break existing functionality}'' (P9). They frequently worked with unfamiliar code written by teammates, requiring extra effort to understand dependencies and integration points (P10). These complexities led developers to favor summaries for building broader system understanding (P8, P9, P11, P13), while also highlighting the importance of summary quality.

\textbf{The absence of LLM knowledge about internal code and high responsibility drove developers to write more detailed prompts and emphasize careful validation.}
LLMs lacked familiarity with domain-specific code and internal infrastructure, making effective prompting more challenging. As P13 noted, ``\textit{Unlike public packages, LLMs don't have this knowledge.}'' Developers compensated by crafting more detailed prompts and manually verifying outputs to ensure correctness. Accountability pressures reinforced this caution, ``\textit{The problems that arise in the industry will ultimately be traced back to the owner of the pull request}'' (P13).

\begin{mdframed}[style=KeyFinding]
\textbf{Key findings:} Developers' prompting choices were shaped by intentional factors. Task urgency favored direct instruction prompting for its speed and flexibility, while goals such as maintainability, learning, and bug avoidance led developers to prefer summary-mediated prompting to support deeper understanding and control. Situational factors further shaped strategy selection: developers preferred direct prompting when tasks were familiar, simple, broad in scope, or clearly defined, and turned to summaries when facing unfamiliar, complex, or ambiguous scenarios. Finally, working with industrial codebases amplified the need for understanding and validation due to large interdependent systems, collaborative workflows, and LLMs' limited knowledge of internal infrastructure.
\end{mdframed}

\subsection{Usability Considerations and Design Opportunities}
\label{sec:usability}

In this section, we examine developers' experiences with the usability of summary-mediated prompting and outline opportunities for design improvement.

\subsubsection{Granularity and Goal Alignment Shape the Usability of Summaries}

\textbf{Usability depends on summary granularity and alignment with developers’ modification goals} (P4, P5, P6, P15).
Participants noted that overly high-level summaries made it difficult to locate precise edit points. As P1 remarked, ``\textit{It's sometimes challenging to determine where to insert or modify my changes [into the summary]}.'' P14 similarly wanted summaries ``\textit{divided into smaller parts, detailing what specific lines do.}'' Conversely, overly fine-grained summaries risked fragmented edits. P6 warned, ``\textit{If A appears multiple times, and you only edit part of it, leftover fragments may remain},'' highlighting the need to balance abstraction and detail to support targeted edits without introducing inconsistencies.

When summaries exactly covered the relevant code and aligned well with developers' goals, participants could make localized edits by simply modifying existing sentences (see Fig.~\ref{fig:prompt_example}). Otherwise, they had to append entire new sentences (P3) or fall back to direct prompting for efficiency (P12).

\subsubsection{Summary-Mediated Prompting Needs Improved Structure, Granularity, Traceability, and Consistency}
\label{sec:design_opportunities}

Developers suggested ways to improve the usability of summary-mediated prompting and better align it with their workflows.

\textbf{Structured formats could better reflect code organization and improve readability.}
Participants preferred structured outlines or bullet points over free-form text to improve readability and reduce cognitive effort (P6, P8, P9, P10, P11). As P8 noted, ``\textit{When the summary is formatted like 1, 2, 3, it becomes easier to read and understand.}'' P9 added that summaries structured by module or function would make writing prompts ``\textit{feel like filling in the blanks.}''

\textbf{Summaries at different granularities were seen as essential for balancing efficiency and thoroughness.}
Participants suggested summaries with adjustable levels of detail, allowing them to start simple and progressively expand as needed (P5, P6, P14). P5 envisioned beginning with a concise version and ``\textit{having a way to expand the summary to include more details.}'' This flexibility would help developers dynamically balance brevity and depth across tasks. It also aligns with \textit{abstraction gradient} in \textit{Cognitive Dimensions of Notations}~\cite{green1989cognitive}, as developers benefit from engaging with representations at varying levels of abstraction depending on their goals.

\textbf{Developers emphasized improving traceability between summaries and code to support comprehension.}
Participants proposed visualizing mappings between summary sentences and corresponding code segments to accelerate understanding (P1, P7). P1 suggested, ``\textit{If I hover over a sentence in the summary, it could show me which part of the code it corresponds to.}'' P13 further recommended highlighting summary portions that require modification to guide developers' attention. These suggestions exemplify \textit{closeness of mapping} and \textit{visibility} dimensions~\cite{green1989cognitive}: making summary–code links explicit helps developers locate relevant context and reduce the mental effort of bridging representations.

\textbf{Consistency in summary generation was critical to support iterative modifications.}
Participants expressed confusion about variability in LLM summaries across iterations. As P4 noted, ``\textit{Each time the summary is generated, the content is different},'' which complicates iterative editing, a common workflow in LLM-assisted programming. Inconsistent outputs disrupted developers' expectations and made it harder to identify what changed and what remained stable, undermining both \textit{consistency} and the \textit{role expressiveness} of generated summaries~\cite{green1989cognitive}. Maintaining consistent summaries that only reflect the modified parts, just like the edited summary in Fig. \ref{fig:prompt_example}, would help preserve developers' mental models over time.

\begin{mdframed}[style=KeyFinding]
\textbf{Key findings:} Developers emphasized that the usability of summary-mediated prompting hinged on both granularity and alignment with modification goals. Summaries that were too high-level obscured edit locations, while overly fine-grained ones introduced inconsistencies. To address these issues, participants suggested concrete improvements: adopting structured formats to reflect code organization, supporting adjustable levels of detail, enabling visual mappings between summaries and code, and maintaining output consistency across iterations.
\end{mdframed}

\section{Threats to Validity}

Our study faces several validity threats. First, the tasks were lightweight to fit within the time constraints of a lab study~\cite{ko2015practical}. Although they span diverse domains and modification types, they do not fully reflect the complexity, scale, and ambiguity of real-world programming. This tradeoff between experimental control and ecological realism is well-documented in programmer user studies~\cite{siegmund2015views, sjoberg2002conducting}.

Second, to focus on prompting strategies rather than bug discovery, we provided explicit modification requirements. While this improved task consistency, it may have affected natural prompting behavior, as participants responded to given goals instead of identifying changes themselves. To reduce this effect, we made the instructions intentionally indirect and prohibited copy-pasting from the instruction document to encourage more active engagement. We also asked participants to avoid manual edits and use prompting whenever possible. These constraints, while necessary for isolating prompting behaviors, may have led participants to write prompts in cases where they would typically make direct edits~\cite{davis2023s}.

Third, differences in the development environment may have influenced participants' behavior. Some were more familiar with macOS and VSCode, while the study used JetBrains IDE on a Windows machine, potentially increasing cognitive load. Running the study via Zoom also introduced a slight screen delay, which may have affected navigation and typing. These factors likely did not affect prompting strategies but may have introduced minor inefficiencies during task execution.

Finally, while our prototype was designed to resemble real-world LLM-assisted coding tools (Section~\ref{sec:design_decisions}), its interface choices (e.g., selection-based prompting) may not reflect the full range of interactions found in AI-powered IDEs. In particular, summary-mediated prompting depends on the capabilities of the underlying LLM, which may shape how developers perceive and engage with generated summaries. Additionally, we relied on self-reported data (e.g., interviews and usability ratings), which are inherently subjective. To mitigate this, we triangulated findings with interaction logs and observations.

\section{Discussion}

\subsection{Cognitive Burden of Prompting for Code Modification}

Prompting for code modification is cognitively demanding, as developers must not only understand the existing code but also communicate their intended changes with clarity and precision. While code summaries offer partial support by facilitating comprehension and providing a structured scaffold, many challenges persist. Developer intentions are often vague or evolving (P7, P10, P13), and relevant context, such as call graphs or dependency structures, is difficult to express precisely in natural language (P9, P15). Although recent vibe coding tools aim to reduce prompting effort by retrieving related code and diagnostic signals, such as linter warnings and console outputs, participants in our study identified a key limitation: LLMs lack visibility into runtime program behavior. Developers found it difficult to describe issues such as unintended UI behavior (P8) or tensor shape mismatches (P14), which only emerge during execution and are hard to convey through prompts. These limitations reflect a persistent gap between what developers observe and what LLMs can interpret, highlighting the need for interaction methods that more effectively capture runtime feedback and evolving intent.

\subsection{Comprehension Challenges in Vibe Coding Workflows}

After committing prompts, developers must comprehend the resulting edits. This is particularly challenging in vibe coding, where LLMs generate or modify large amounts of code in one step, often across multiple files, leaving developers overwhelmed and unsure of what changed. Tools like Bolt\footnote{\url{https://bolt.new/}} exemplify this trend, aiming to support low- or no-code development through conversational prompts and runtime previews. While many developers appreciated the efficiency, our findings indicate that comprehension remains essential, particularly for those prioritizing maintainability, learning, or bug avoidance. Several participants preferred incremental updates for greater control (P10) and expressed low trust in large or cross-file modifications (P3, P5, P7, P8, P13), highlighting the difficulty of maintaining trust and oversight in current vibe coding tools. Basic code diffs and chatbot-style explanations may be insufficient for understanding and validating large-scale LLM edits, particularly for end users with limited technical background. Future studies should examine this beyond small-snippet completions (e.g., Copilot~\cite{tang2024developer}) and across different user groups. Our findings suggest that structured summaries and code–summary mappings can scaffold comprehension and align edits with developer goals. Future work should also explore integrating traditional techniques, such as automated testing, to enhance trust in multi-step modifications.

\subsection{Natural Language Programming in the LLM Era}

Natural language programming aims to let people express ideas ``in the same way they think about them''~\cite{myers2004natural}, offering a more intuitive bridge between mental intent and computational logic. While early advocates saw it as a path to democratizing programming~\cite{sammet1966use}, critics argued that natural language, with its inherent ambiguity, lacks the precision required for instructing machines~\cite{dijkstra2005foolishness}. Modern LLMs represent a shift: rather than replacing programming languages, they can interpret vague or underspecified intent, positioning natural language as an interactive layer between human reasoning and formal code. Our study shows that treating summaries as a medium for intent scaffolding reveals several key usability factors. Further research should explore the broader roles and design spaces of natural language across diverse programming contexts.

\section{Conclusion}

This study examined how developers construct prompts for LLM-assisted code modification, comparing direct instruction and summary-mediated prompting. Through a mixed-methods study with 15 developers across diverse programming tasks, we characterized their prompting workflows and decision-making patterns. Building on these findings, future work should explore interactive systems that better scaffold comprehension and control through code summaries designed for structural clarity, improved traceability, and adjustable granularity. In addition, longitudinal deployment studies are needed to examine how prompting strategies evolve over time and integrate into real-world development workflows at scale.

\section*{Data and Code Availability}

To support transparency and reproducibility, we provide a replication package with the codebook, coded interview segments, interaction logs, questionnaire responses, analysis scripts, study protocol, task descriptions, and source code of \system, available at: 
\url{https://github.com/ND-SaNDwichLAB/direct-vs-summary-study}.

\section*{Acknowledgment}
This research was supported in part by an AnalytiXIN Faculty Fellowship, an NVIDIA Academic Hardware Grant, a Google Cloud Research Credit Award, a Google Research Scholar Award, and NSF grants CCF-2211428, CCF-2315887, and CCF-2100035. Any opinions, findings, or recommendations expressed here are those of the authors and do not necessarily reflect the views of the sponsors.

\bibliographystyle{IEEEtran}
\bibliography{reference}

\begin{threeparttable}
\centering
\caption{Summary of Participant Demographics and Technical Background.}
\footnotesize
\label{tab:demographics}
\begin{tabular}{c|cccc|ccc|ccc}
\toprule
ID & Gender & Age & Role & Experience & Deep Learning & Data Vis & Web Dev & TensorFlow & D3.js & Chrome Ext \\
\midrule \midrule
P1 & Male & 27 & Graduate Student & 8 years & Yes & Yes & Yes & Proficient & Proficient & Proficient \\
P2 & Male & 22 & Graduate Student & 5 years & Yes & Yes & Yes & Unfamiliar & Never & Never \\
P3 & Male & 24 & Graduate Student & 6 years & Yes & Yes & Yes & Never & Never & Never \\
P4 & Male & 22 & Graduate Student & 5 years & Yes & Yes & Yes & Unfamiliar & Unfamiliar & Unfamiliar \\
P5 & Female & 27 & Graduate Student & 10 years & Yes & Yes & Yes & Unfamiliar & Unfamiliar & Never \\
P6 & Female & 23 & Graduate Student & 6 years & Yes & Yes & Yes & Unfamiliar & Never & Proficient \\
P7 & Male & 27 & Graduate Student & 7 years & Yes & Yes & Yes & Never & Never & Unfamiliar \\
P8 & Male & 25 & Professional Developer & 6 years & No & Yes & Yes & Never & Never & Proficient \\
P9 & Male & 24 & Professional Developer & 7 years & No & Yes & Yes & Unfamiliar & Never & Never \\
P10 & Male & 22 & Professional Developer & 6 years & Yes & Yes & Yes & Proficient & Unfamiliar & Never \\
P11 & Male & 28 & Professional Developer & 8 years & Yes & Yes & No & Proficient & Never & Unfamiliar \\
P12 & Female & 27 & Graduate Student & 5 years & Yes & Yes & Yes & Unfamiliar & Never & Unfamiliar \\
P13 & Male & 30 & Professional Developer & 12 years & Yes & Yes & Yes & Unfamiliar & Unfamiliar & Unfamiliar \\
P14 & Female & 22 & Graduate Student & 5 years & Yes & No & Yes & Unfamiliar & Never & Unfamiliar \\
P15 & Female & 27 & Graduate Student & 8 years & Yes & Yes & Yes & Unfamiliar & Unfamiliar & Unfamiliar \\
\bottomrule
\end{tabular}
\begin{tablenotes}
\footnotesize
\item[*] Data Vis = Data Visualization; Web Dev = Web Development; Chrome Ext = Chrome Extension Development.
\item[†] Deep Learning, Data Vis, and Web Dev indicate whether the participant had prior experience in each domain.
\item[‡] TensorFlow, D3.js, and Chrome Ext indicate their self-rated proficiency levels.
\end{tablenotes}
\end{threeparttable}

\end{document}